\documentstyle[twocolumn,psfig,aps,amsfonts,amssymb,prl]{revtex}

\begin{document}
\draft
\preprint{{\bf ETH-TH/98-04}}

\title{Supercurrent Quantization in Narrow Channel SNS Junctions}

\author{Nikolai M.\ Chtchelkatchev$^{a\,}$, Gordey B.\ Lesovik$^{a\,}$ 
and Gianni Blatter$^{b\,}$}

\address{$^{a\,}$L. D. Landau Institute for Theoretical Physics,
  117940 Moscow, Russia}

\address{$^{b\,}$Theoretische Physik, ETH-H\"onggerberg, CH-8093
  Z\"urich, Switzerland}

\twocolumn[
\date{\today}
\maketitle
\widetext
\vspace*{-1.0truecm}
\begin{abstract}
\begin{center}
\parbox{14cm}{ We determine the quasi-particle excitation spectrum in
the normal region of a narrow ballistic
superconductor--normal-metal--superconductor (SNS) Josephson
contact. Increasing the effective chemical potential in the contact
converts the electronic levels into Andreev-levels carrying
supercurrent. The opening of these superchannels leads to a
supercurrent quantization which exhibits a non-universal behavior in
general and we discuss its dependence on the junction parameters.}
\end{center}
\end{abstract}
]
%\pacs{PACS numbers: 74.80.Fp, 74.50.+r, 74.60.Jg}

\vspace{-0.4truecm}
\narrowtext

The Josephson effect \cite{Josephson}, a hallmark of
superconductivity, is of fundamental interest and bears considerable
potential for applications in superconducting electronics
\cite{Likharev}.  Today, the miniaturization of electronic structures
has reached the regime where the transport proceeds via few or even a
single conducting channel \cite{vanWees}. Using gated structures,
junctions can be transformed from insulating SIS to superconducting
SNS states with the quasi-particle spectrum evolving from
phase-insensitive electronic- to phase-sensitive Andreev states
carrying large supercurrents.  The onset of superflow proceeds in
steps associated with the opening of transverse channels, similar to
the conductance quantization in a metallic contact. While in the
latter the conductance $G = I/V$ is quantized in units of $2e^2/\hbar$
\cite{vanWees,Glazman}, it is the maximal supercurrent $I_c$ which is
quantized in units of $e \Delta/\hbar$ in the case of a short
Josephson link \cite{BeenakkervanHouten,Furusaki} ($\Delta$ is the
superconducting gap in the banks).  However, contrary to the
universality of the quantization in a normal contact, the quantization
of the critical supercurrent is perfect only in the limit of short
junctions $L\ll\xi_0$ ($\xi_0=\hbar v_{\rm\scriptscriptstyle
F}/\pi\Delta$ is the superconducting coherence length) but is
non-universal in general \cite{Furusaki}. Indeed, while experiments on
superconducting quantum point contacts do show steps in the critical
current $I_c$, these are much less prominent than the corresponding
steps in the conductance $G$ \cite{Takayanagi}.  In this letter, we
study the opening of superconducting channels in the metallic link of
a narrow ballistic SNS Josephson contact and determine the evolution
of the quasi-particle spectrum and the dependence of the supercurrent
quantization on the junction parameters.

While the behavior of macroscopic SNS Josephson junctions is well
understood \cite{KulikIshii}, the present interest concentrates on
gated structures of mesoscopic size. Such quantum point contacts are
realized in heterostructures \cite{vanWees,Takayanagi} or via
manipulations with a scanning tunneling microscope \cite{STM}.
Theoretically, the supercurrent-phase relation in mesoscopic SNS
junctions with a $\delta$-scatterer has been analyzed \cite{Bagwell}
and the phenomenon of supercurrent quantization has been studied in
short junctions \cite{BeenakkervanHouten}. Non-universal features of
supercurrent quantization have first been observed by Furusaki {\it et
al.} \cite{Furusaki} --- unfortunately, these numerical results
provide limited insight into the physical origin and the parametric
dependence of these effects.  Here, we present a detailed discussion
of the opening of superchannels in mesoscopic SNS junctions using
quasi-classical and scattering matrix techniques. We discuss the
non-trivial evolution of the excitation spectrum as the chemical
potential drops below the superconducting gap and analyze the
transformation of the ballistic SNS structure into a SIS tunnel
junction. The phase-dependence of the quasi-particle spectrum allows
us to find the supercurrent quantization in short and long junctions,
where the contribution from the continuous spectrum can be ignored.
\begin{figure} [bt]
\centerline{\psfig{file=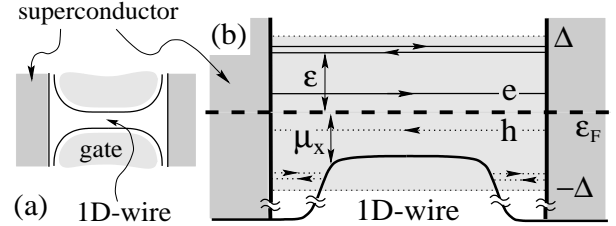,width=8.0cm,height=3.0cm}}
\narrowtext\vspace{4mm}
\caption{Narrow channel SNS contact: (a) geometrical setup showing the
adiabatic joining of the wire to the superconductor, (b) potential
landscape with a flat potential in the wire center and smoothly
dropping to the band bottom in the superconducting banks. While
electrons and holes with small excitation energies $\varepsilon<\mu_x$
form current carrying Andreev states, the hole propagation is quenched
at large energies $\epsilon > \mu_x$ and the Andreev levels transform
into localized electronic states.}
\end{figure}
Consider a narrow metallic lead (with few transverse channels)
connecting two superconducting contacts as sketched in Fig.\ 1(a) [we
assume piecewise constant gap parameters $\Delta (x<-L/2)=$
$\Delta\exp(i\varphi_{\rm \scriptscriptstyle L})$,
$\Delta(|x|<L/2)=0$, and $\Delta(L/2<x)= \Delta$
$\exp(i\varphi_{\rm\scriptscriptstyle R})$]. For each transverse
channel in the metallic wire the quasi-particle spectrum
$\varepsilon_\nu$ is determined through the 1D Bogoliubov-de Gennes
equation (we choose states with $\varepsilon_\nu \geq 0$)
\[
  \left[ \begin{array}{cc}
  \!-\frac{\hbar^2 \partial_x^2}{2m} - \mu_x(x) & \!\! \Delta(x) \! \\
  \!\! \Delta^*(x) & \!\!\frac{\hbar^2\partial_x^2}{2m}+\mu_x(x) \!
  \end{array}\right]
  \! \left[\begin{array}{c}\! u_\nu(x)\! \\
  \! v_\nu(x) \! \end{array} \right] = \varepsilon_\nu
  \! \left[\begin{array}{c}\! u_\nu(x)\! \\
  \! v_\nu(x) \! \end{array} \right],
\]
where $u_\nu$ and $v_\nu$ denote the electron- and hole-like
components of the wave function $\Psi_\nu$. The spectrum splits into
continuous and discrete parts and we concentrate on the latter part in
the following, $\varepsilon_\nu < \Delta$.

The effective chemical potential $\mu_x(x) =
\varepsilon_{\rm\scriptscriptstyle F} - \varepsilon_\perp(x)$ accounts
for the transverse energy $\varepsilon_\perp(x)$ of the channel, see
Fig.\ 1(b).  Within a quasi-classical formulation we describe the
quasi-particles in terms of their kinetic energies $K_\pm = \hbar^2
k_\pm^2/2m = \mu_x(x) \pm \varepsilon$ and assume transmission and
reflection to be ideal (the excitation energies $\varepsilon = E
-\varepsilon_{\rm\scriptscriptstyle F} > 0$ are measured with respect
to the Fermi energy $\varepsilon_{\rm\scriptscriptstyle F}$). An
electron with energy $\varepsilon < \Delta$ below the gap is reflected
back from the superconductor as a hole with kinetic energy
$K_-=\mu_x-\varepsilon$, injecting a Cooper-pair into the
superconducting contact, a process known as Andreev reflection
\cite{Andreev}. A second reflection at the opposite NS boundary
transforms the hole state back into the original electron state, thus
producing a phase sensitive Andreev level carrying the supercurrent
across the normal metal lead. The hole-part associated with the
Andreev level can propagate only if its kinetic energy is positive,
$K_- >0$, see Fig.\ 1(b). Otherwise, the hole is back-reflected from
the normal potential in the junction and transformed into an electron
at the NS boundary --- the incident electron is effectively reflected
back as an electron and a phase-insensitive electronic level is
formed.  Hence, the superchannel starts being modified when the
chemical potential $\mu_x$ drops below the gap $\Delta$ and is
quenched completely with all Andreev levels transformed into
electronic ones when $\mu_x$ becomes negative.

Going beyond quasi-classics, the above physics is conveniently
described through the scattering matrix formalism
\cite{BuettikerLandauer,Beenakker}.  We define scattering states in
the normal region and characterize them through the energy dependent
transmission and reflection coefficients $t\exp(i\chi^t)$ and
$r\exp(i\chi^r)$ describing the propagation of quasi-particles
incident from the left through the junction.  Matching these states
with the evanescent modes in the superconductors we obtain (within the
Andreev approximation \cite{Andreev}: $(K_+ -K_-)/(K_+ +K_-) \ll 1$ at
the NS interface) the quantization condition,
\begin{equation}
\cos(S_{+}-S_{-}-\alpha)=r_{+}r_{-}\cos\beta+t_{+}t_{-}\cos\varphi,
\label{scat}
\end{equation}
where the $+(-)$ signs refer to the positive and negative energies
$\pm \varepsilon$ of the electron(hole)-like quasi-particles and
$S_\pm(\varepsilon)=\chi^t_\pm+k_{0,\pm} L$, with $k_{0,\pm} L=
\sqrt{2m(\varepsilon_{\rm\scriptscriptstyle F}\pm \varepsilon)}L/
\hbar$ the phase for free propagation (while the phase $S$ refers to
the propagation from $-L/2$ to $L/2$, the scattering phases $\chi^t$
and $\chi^r$ refer to the origin). Andreev scattering at the NS
boundaries introduces the phase $\alpha = 2 {\rm arccos}
(\varepsilon/\Delta)$ ranging from $\pi$ at $\varepsilon = 0$ to 0 at
the gap $\varepsilon = \Delta$, as well as the phase difference
$\varphi = \varphi_{\rm\scriptscriptstyle L} -
\varphi_{\rm\scriptscriptstyle R}$ between the two superconducting
banks. The phase $\beta = (\chi^t_+ - \chi^r_+) - (\chi^t_- -
\chi^r_-)$ reduces to $\beta = 0$ for a symmetric barrier in the
absence of perfect resonances [as follows from the unitarity of the
scattering matrix; for a symmetric potential shifted by $a$ from the
center we have $\beta=2(k_{0,+}\!-\!k_{0,-})a$]. {The secular equation
(\ref{scat}) involves two main energy dependencies originating from
the propagation through the normal layer} ($S\pm$) and from scattering
at the NS boundaries, e.g., due to potential steps or barriers.  Here,
we concentrate on the case where the transport through the junction is
dominated by the normal metallic wire --- we will comment on the
effect of resonances introduced by an additional scattering at the NS
boundaries below.
\begin{figure} [bt]
\centerline{\psfig{file=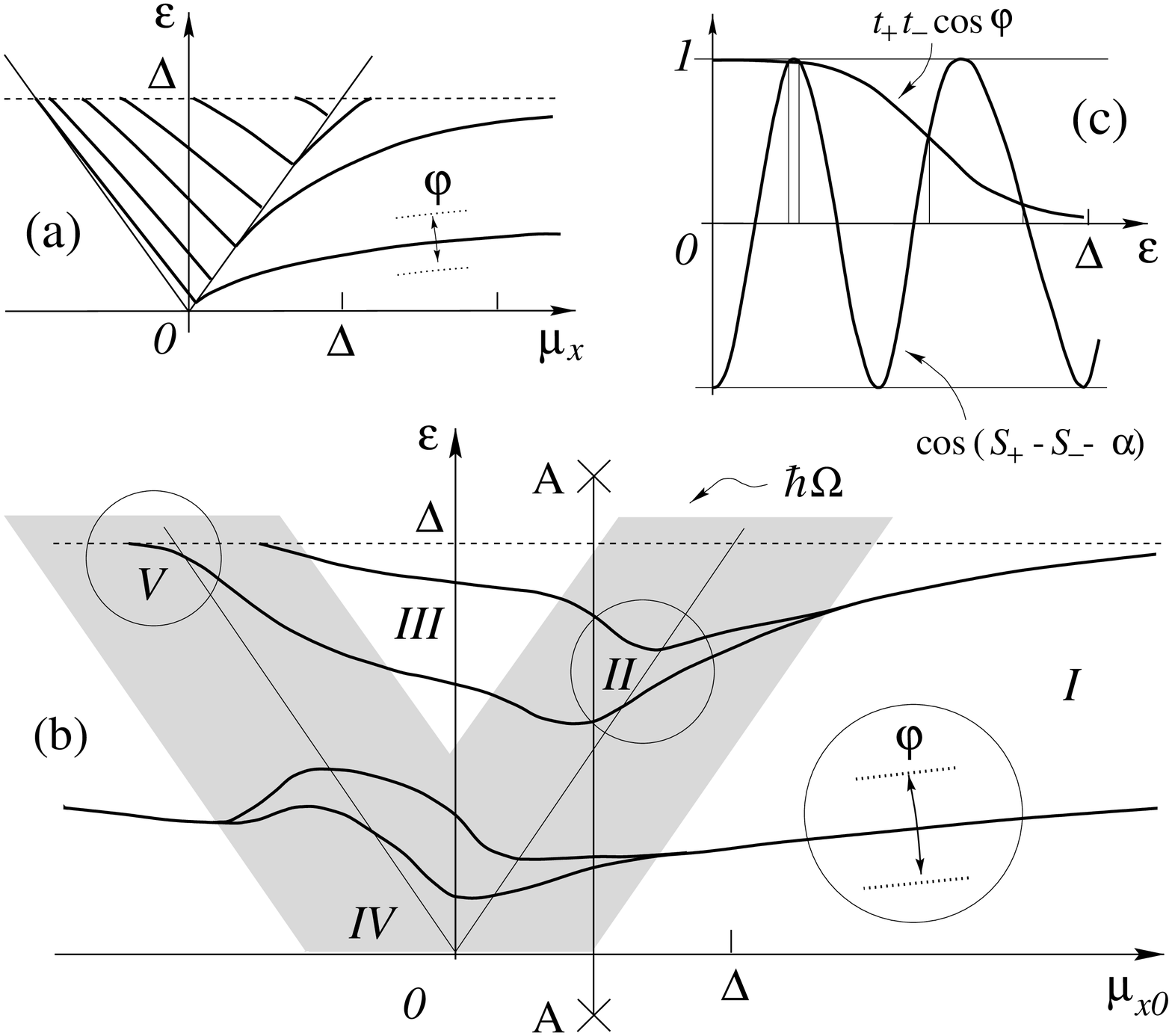,width=7cm,height=5.9cm}}
\narrowtext\vspace{4mm}
\caption{Discrete energy spectrum [(a) qualitative sketch for a flat
potential, (b) smooth parabolic potential]: For $\mu_{x0} >
\varepsilon$ (region I) both electrons and holes propagate, forming
phase ($\varphi$) sensitive Andreev levels carrying supercurrent. The
double degeneracy of the Andreev states is lifted by a finite phase
drop $\varphi$ across the junction as well as a finite reflection in
the wire (see (c)), the latter becoming relevant upon decreasing
$\mu_{x0}$.  As $\mu_{x0}$ drops below $\varepsilon$ the Andreev
levels first transform into electronic states (regions II and III) and
finally turn into boundary states trapped at the NS interfaces when
$\mu_{x0} \alt -\varepsilon$ (region V). Within the shaded regions
around $\varepsilon = \pm \mu_{x0}$ the transmission drops from unity
to zero for holes and electrons. (c): graphical solution of (1) along
the cut A-A in (b).}
\end{figure}
A rough understanding of the transformation from a metallic to an
insulating junction is obtained in the quasi-classical approximation
using a flat potential, see Fig.\ 1(b): For a large chemical potential
$\mu_x> \varepsilon$ we have $r_\pm = 0$, $t_\pm = 1$ in (\ref{scat})
and we obtain the Bohr-Sommerfeld quantization condition for the
(phase sensitive) Andreev levels $S_+ - S_- -\alpha \pm \varphi = 2 n
\pi$.  Evaluating this condition for a flat potential (ignoring
contributions to $S_\pm = k_{{\rm\scriptscriptstyle F},x} L \sqrt{1\pm
\varepsilon/\mu_x}$ originating from the adiabatic joints) we obtain
the level scheme shown in Fig.\ 2(a).  On the other hand, for
$-\varepsilon < \mu_x < \varepsilon$ the right hand side of
(\ref{scat}) vanishes and using $S_- = 3\pi/2$ (assuming a hard wall
potential) we find the quantization condition $2S_+ -2 \alpha = 2 n
\pi$, for the electronic levels, see Fig.\ 2(a) for a qualitative
result in a flat potential. Note that we have twice as many electronic
than Andreev levels as the latter are doubly degenerate at $\varphi =
0$ --- the exact transformation of the Andreev levels in electronic
ones at $\varepsilon \approx \mu_x$ requires a more careful analysis
accounting for the non-ideal transmission and reflection through the
normal channel, see below. Finally, as $\mu_x$ drops below
$-\varepsilon$ both the electron- and hole-like trajectories are
quenched. Note the evolution of the r.h.s.\ of (\ref{scat}), going
from $\cos \varphi$ at large positive $\mu_x > \varepsilon$, to a
small value in the intermediate region $-\varepsilon < \mu <
\varepsilon$, and back to unity at large negative $\mu_x <
-\varepsilon$. This provides us with a first rough understanding of
the SNS to SIS transformation.

In a more accurate study of the evolution of the bound state spectrum
from a SNS to a SIS junction we assume a smooth potential $\mu_x(x)$
with a small curvature $m\Omega^2\!=\!\partial_x^2\mu_x$,
$\hbar\Omega\!<\!\Delta$, producing a sharp switching between
transmission and reflection within the energy interval $\hbar \Omega$
(a $\delta$-function scatterer \cite{Bagwell} does not describe a
pronounced transformation from a SNS to a SIS junction). Adiabatic
joining of the wire to the superconducting banks requires that
$m\Omega^2(L/2)^2/2 \sim \varepsilon_{\rm\scriptscriptstyle F}$ and
allows us to make use of the Andreev approximation while avoiding the
appearance of resonances (this condition can be relaxed as the Andreev
approximation requires $m\Omega^2 (L/2)^2/2 \! \gg \!  \Delta$, while
a step in the potential $\Delta V \!<\! 0.9 \varepsilon_{\rm
\scriptscriptstyle F}$ produces only weak resonances]. In summary, a
smooth contact without resonances requires the parameter setting
$\sqrt{\varepsilon_{\rm\scriptscriptstyle F}
\varepsilon_{\rm\scriptscriptstyle L}}\! < \!\hbar \Omega \!<\!
\Delta$ where $\varepsilon_{\scriptscriptstyle L} \! \equiv$ $\hbar^2
\pi^2/ 2 m L^2$; this condition implies a long junction $L > \xi_0$
and hence a relatively large number $n \sim \sqrt {\varepsilon_{\rm
\scriptscriptstyle F}/\Delta}$ of trapped levels. For such a smooth
potential the Kemble formula is valid and the transmission
probabilities take the form $T_\pm = t_\pm^2 =
1/\{1+\exp[-2\pi(\mu_x(0)\pm \varepsilon)/\hbar\Omega]\}$
\cite{Glazman}.

Fig.\ 2(b) shows the refined results for the SNS to SIS transformation
using the quadratic potential $\mu_x(x)=\mu_{x0} + m \Omega^2 x^2/2$
with the parameter $\hbar \Omega(\mu_{x0}) =(4/\pi)
\sqrt{\varepsilon_{\scriptscriptstyle L}
(\varepsilon_{\rm\scriptscriptstyle F}-\mu_{x0})}$ and $\mu_{x0} =
\mu_x(0)$, joining the band bottom of the superconductors at the two
NS boundaries. For this case, the quasi-classical dimensionless action
takes the form $S(E)/\hbar = (2E/\hbar\Omega)[\kappa^2
\sqrt{1+\kappa^{-2}} +\ln[|\kappa|(1+\sqrt{1+\kappa^{-2}})]$ with
$\kappa^2 = Q\,\hbar\Omega/E = \pi^2 \hbar^2\Omega^2/ 16 E
\varepsilon_{\scriptscriptstyle L}$, $Q \gg 1$ a large parameter
[$S_\pm = S(E=\mu_{x0} \pm \varepsilon)$; an additional phase $\pi$,
which cannot be obtained within the quasi-classical scheme, is picked
up over the energy interval $\hbar\Omega$ as $E$ goes through zero].
As for the flat potential, the Andreev levels at large chemical
potential $\mu_{x0} > \varepsilon + \hbar\Omega$ (region I) are
converted in steps (regions II --- V) to the electronic states at
negative potential $\mu_{x0} < -\varepsilon - \hbar\Omega$: Upon
entering region II the product $t_{+}t_{-}$ in (\ref{scat}) drops
below unity and the Andreev levels split even for $\varphi \!=\!
0$. In region III the hole propagation is quenched and electronic
levels with an exponentially weak phase sensitivity are
formed. Entering region V the electronic propagation through the wire
is suppressed and pairs of boundary states are formed near the NS
interfaces \cite{WendinSchumeiko} (for the flat potential these
boundary states are lifted to $\varepsilon \approx \Delta$ as the
Andreev phase $\alpha$ has to vanish).  In region IV both the
electron- and hole-components undergo finite reflection and the
distinction between Andreev- and electronic levels is gone; this
resembles the situation of a SNS junction with a $\delta$-scattering
potential \cite{Bagwell}. More details of this conversion between
Andreev- and electronic levels will be given elsewhere
\cite{Kuhn}. Below, we concentrate on the quasi-classical region I and
study the evolution of the cri\-tical supercurrent as the channel is
switched on and off.

The supercurrent $I$ flowing through the junction splits into the two
contributions from the discrete $(I_{\rm dis})$ and the continuous
$(I_{\rm con})$ parts of the spectrum. Here, we concentrate on $I_{\rm
dis}$, which dominates the expression for the critical supercurrent in
the quasi-classical regime I (requiring $\hbar\Omega < \mu_{x0}$ is
sufficient). The current of the $\nu$-th level (including a factor 2
for spin) can be obtained from the derivative $I_\nu = (2e/\hbar)
\partial_\varphi\varepsilon_\nu = (2e/{\cal T}) \, t_+t_- \sin
\varphi$, with the generalized travelling time ${\cal T}=\sin(\delta
S-\alpha) \hbar\partial_\varepsilon[\delta S-\alpha]$
$+\hbar\partial_\varepsilon [(t_+t_-)\cos\varphi
+(r_+r_-)\cos\beta]$. Within the quasi-classical region I, each
Andreev level carries a finite supercurrent of amplitude
$2e/[\tau_++\tau_-+2\hbar/\sqrt{\Delta^2- \varepsilon^2}]$, where
$\tau_\pm = \hbar \partial_\varepsilon S_\pm$ denote the propagation
times for the electron- and hole-components; for the smooth quadratic
potential we find $\tau(E)=\Omega^{-1}
[2\ln[|\kappa|(1+\sqrt{1+\kappa^{-2}})]$.  For small energies the
travel time increases logarithmically $\tau (E) \approx \Omega^{-1}
\ln(4Q\hbar\Omega/E)$ within the interval $\hbar \Omega < E < Q
\hbar\Omega$ and saturates at $\tau_0 \approx \Omega^{-1} \ln(4Q)$ as
$E$ drops below $\hbar \Omega$, a result going beyond the
quasi-classical approximation.  At $\varphi = 0$, the pairwise
degenerate levels produce equal currents of opposite sign and the sum
over the discrete spectrum gives no current. Increasing $\varphi$, the
degeneracy is lifted and the resulting miscancellation leads to a
finite supercurrent.  Each pair produces a monotonuously growing
current of the same sign, hence the largest current is reached at
$\varphi = \pi^-$. However, at $\varphi = \pi^-$ the levels become
degenerate again and their currents cancel pairwise, except for the
lowest level which remains unpaired and thus carries all the
supercurrent from the discrete part of the spectrum. The continuous
part of the spectrum vanishes at $\varphi = \pi$, however, this is
{\it a priori} not sufficient to guarantee that the critical current
$I_c$ is the current $I_0$ carried by the lowest level --- we have to
show in addition that the maximum of $I = I_{\rm dis} + I_{\rm con}$
is reached at $\varphi = \pi^-$ [indeed, we could prove that this
condition is fulfilled within a regime of the $L$-$\mu_{x0}$ plane
away from $(L \sim (\xi_0/ k_{\rm\scriptscriptstyle F})^{1/2},\mu_{x0}
\sim \Delta)$]. In the end, we arrive at a particularly simple
expression for the critical current density in the quasi-classical
region I,
\begin{equation}
I_c = e/(\tau_0 + \hbar/\Delta).
\label{crit_current}
\end{equation}
The travel time $\tau_0$ is constant ($\Omega^{-1}\ln 4Q$) at the
opening of the channel, decreases as $\Omega^{-1}\ln
4Q\hbar\Omega/\mu_{x0}$ for $\mu_{x0} > \hbar\Omega$ and transforms to
the free travel time $L/v_{{\rm\scriptscriptstyle F},x}$ for $\mu_{x0}
> Q \hbar\Omega$.  As the channel becomes wide open at high energies,
the critical current saturates to the expected value $I_c =
ev_{\rm\scriptscriptstyle F} /(L+\pi\xi_0)$.

The above discussion dealt with the parameter settings
$\sqrt{\varepsilon_{\rm\scriptscriptstyle F}
\varepsilon_{\rm\scriptscriptstyle L}} < \hbar \Omega < \Delta$
requiring a junction with $L > \xi_0$.  Releasing the condition of
small curvature and assuming $\Delta < \hbar\Omega$, the SNS to SIS
transformation is smeared and the region IV occupies all of the
interesting crossover regime. For a flat potential $\hbar \Omega <
\sqrt{\varepsilon_{\rm\scriptscriptstyle F}
\varepsilon_{\scriptscriptstyle L}}$ the situation is complicated by
the appearance of resonances due to reflection from the potential step
at the NS boundary \cite{Kuhn}. The situation simplifies for a very
short junction with $L \ll \xi_0$, where we can again make use of
(\ref{scat}) to produce a simple and universal result
\cite{FurusakiTsukuda}: with $\delta S = S_+ - S_- \approx 0$ and $t_-
\approx t_+\approx \sqrt{T}$, $r_-\approx r_+ \approx \sqrt{R}$, we
find that only one level remains trapped in the junction at
$\varepsilon_0=\Delta [1-T\sin^2(\varphi/2)]^{1/2}$.  Here, we require
a width $\hbar \Omega > \Delta$ in order to avoid a strong energy
dependence in the transmission probability $T$.  Determining the
current $I_0(\varphi)$ from $\varepsilon_0$ and maximizing, we obtain
\begin{equation}
I_c = (e\Delta/\hbar)\, (1-\sqrt{R}),
\label{crit_current_short}
\end{equation}
in marked difference from the result for the conductance quantization
$G = (2e^2/\hbar)(1-R)$: a finite reflection $0<R \ll 1$ will affect
the supercurrent quantization much more strongly than the conductance
quantization.
\begin{figure} [bt]
\centerline{\psfig{file=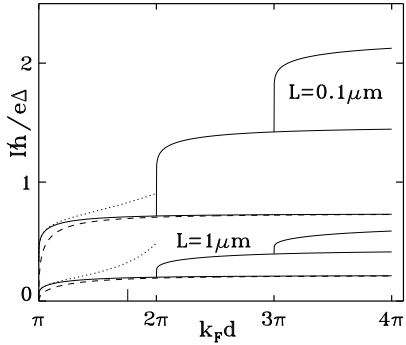,width=6.0cm,height=4.8cm}}
\narrowtext\vspace{4mm}
\caption{Supercurrent quantization: with increasing width $d$ of the
normal channel the supercurrent increases in steps of
$e/(\tau_0+\hbar/\Delta)$. The dotted and dashed lines give the
approximations $\tau_0 \approx \Omega^{-1} \ln (4 Q \hbar \Omega/E)$
and $\tau_0 \approx L/v_{\rm\scriptscriptstyle F}$ at small and large
energies (parameters: $\varepsilon_{\rm\scriptscriptstyle F} = 1$ eV,
$\Delta/\varepsilon_{\rm\scriptscriptstyle F} = 10^{-3}$, $L/\xi_0
\sim 1,\,10$; with the curvature $\hbar\Omega/\Delta < 5,\,0.5$ no
smearing is visible at the supercurrent onset).}
\end{figure}
Finally, we discuss the supercurrent quantization `steps' appearing as
the gate potential is decreased to open the conducting channel. We
concentrate on the quasi-classical regime I, assuming a parabolic
potential $\mu_x(x)$ in the junction which matches the band bottom of
the superconductors at the NS boundary and ignore a possible change in
the effective mass.  The quantized transverse energy of a channel of
width $d$ is given by $\varepsilon_{\perp;l} \approx\hbar^2 \pi^2
l^2/2 m d^2$; these levels match up with the Fermi energy when $d =
d_k = k\pi/k_{\rm\scriptscriptstyle F}$.  As we open the $k$-th
channel the other open channels have already dropped by
$\mu_{x0;l,k}=\varepsilon_{\rm\scriptscriptstyle F}
-\varepsilon_{\perp;l,k} = \varepsilon_{\rm\scriptscriptstyle
F}(1-l^2/k^2)$, where $\varepsilon_{\perp;l,k} = \hbar^2 \pi^2 l^2/2 m
d_k^2$, e.g., the first channel is wide open when the second channel
appears, $\mu_{x0;1,2} = (3/4) \varepsilon_{\rm\scriptscriptstyle F}$.
Increasing the channel width $d$, the first channel opens (i.e.,
$\mu_{x0} = \varepsilon_{\rm\scriptscriptstyle F}
(1-\pi^2/k_{\rm\scriptscriptstyle F}^2d^2)$ turns positive) as we
reach $d_1=\pi/k_{\rm\scriptscriptstyle F}$, the critical current
increases sharply $I_c \approx e \Omega / \ln
[4/(k_{\rm\scriptscriptstyle F}^2 d^2/\pi^2-1)]$ (the logarithmic
singularity is cutoff at $\hbar \Omega$) and saturates at $I_c \approx
e v_{\rm\scriptscriptstyle F}/(L+\pi\xi_0)$, see Fig.\ 3. Here, we
have ignored the smearing near the onset within the range
$\hbar\Omega$ due to a finite reflection --- while the interesting
evolution of the quasi-particle spectrum is washed out as
$\hbar\Omega$ increases beyond the gap $\Delta$, see Fig.\ 2, the
(smooth) steps in the onset of the supercurrent are much more robust.
On the other hand, the absence of sharp steps in the critical current
onset is an intrinsic feature of the superconducting junction.  For
the short junction with $L \ll \xi$ the sharpness of the steps in
$I_c$ is dictated by the reflection probability $R$ of the junction
and thus is more similar to the steps in the conductance $G$. However,
with $I_c \propto 1-\sqrt{R}$, the steps in $I_c$ are always smoother
than those in the conductance $G \propto (1-R)$.

In the end, universal supercurrent quantization first seems to require
short junctions, but the gate needed to switch the channels will
produce backscattering and spoil the quantization. While going over to
longer contacts helps to produce sharp conductance steps, the onset of
supercurrent remains smooth due to the long travelling time for the
Andreev states.

We thank D.\ Kuhn, G.\ Wendin, and V.\ Shumeiko for helpful
discussions and the Swiss National Foundation for financial support.


\begin{thebibliography}{99}
\bibitem{comment} While the condition (\ref{scat}) 
has been presented in a determinantal form \cite{Beenakker},
its potential for the analysis of the SNS to SIS transformation
has not been recognized.
\end{thebibliography}
\end{document}